# Tunable interlayer magnetism and band topology in van der Waals heterostructures of MnBi$_2$Te$_4$-family materials

Zhe Li[1,2,†], Jiaheng Li[1,2,†], Ke He[1,2,3]*, Xiangang Wan[4,5], Wenhui Duan[1,2,3,6], and Yong Xu[1,2,7,*]

[1]*State Key Laboratory of Low-Dimensional Quantum Physics, Department of Physics, Tsinghua University, Beijing 100084, China*

[2]*Frontier Science Center for Quantum Information, Beijing 100084, China*

[3]*Bejing Institute of Quantum Information Science, Beijing 100193, China*

[4]*National Laboratory of Solid State Microstructures and School of Physics, Nanjing University, Nanjing 210093, China*

[5]*Collaborative Innovation Center of Advanced Microstructures, Nanjing 210093, China*

[6]*Institute for Advanced Study, Tsinghua University, Beijing 100084, China*

[7]*RIKEN Center for Emergent Matter Science (CEMS), Wako, Saitama 351-0198, Japan*

[†]These authors contributed equally to this work.

*Correspondence to: yongxu@mail.tsinghua.edu.cn (Y. X.); kehe@tsinghua.edu.cn (K. H.).

**Manipulating the interlayer magnetic coupling in van der Waals magnetic materials and heterostructures is the key to tailoring their magnetic and electronic properties for various electronic applications and fundamental studies in condensed matter physics. By utilizing the MnBi$_2$Te$_4$-family compounds and their heterostructures as a model system, we systematically studied the dependence of the sign and strength of interlayer magnetic coupling on constituent elements by using first-principles calculations. It was found that the coupling is a long-range superexchange interaction mediated by the chains of *p* orbitals between the magnetic atoms of neighboring septuple-layers. The interlayer exchange is always antiferromagnetic in the pure compounds, but can be tuned to ferromagnetic in**

**some combinations of heterostructures, dictated by *d* orbital occupations. Strong interlayer magnetic coupling can be realized if the medial *p* electrons are delocalized and the *d* bands of magnetic atoms are near the Fermi level. The knowledge on the interlayer coupling mechanism enables us to engineer magnetic and topological properties of MnBi$_2$Te$_4$-family materials as well as many other insulating van der Waals magnetic materials and heterostructures.**

Van der Waals (vdW) layered magnets have attracted intensive attentions in the recent years (*1-6*). Benefitting from the weak vdW interlayer bonding, these materials can be easily exfoliated into ultrathin flakes and flexibly recombined into various heterostructures with novel material properties. Many layered magnetic materials have been found, including Cr$_2$Ge$_2$Te$_6$, CrI$_3$, Fe$_2$Ge$_2$Te$_6$, FePS$_3$, VSe$_2$, MnSe$_2$ (*7-22*), and the MnBi$_2$Te$_4$-family compounds (*23-34*), which provide new chances for developing electronics, spintronics, optoelectronics, and topological physics (*4-6, 23-39*). A key ingredient to designing and engineering vdW magnetic materials and heterostructures for various purposes is to manipulate the interlayer magnetic coupling (IMC), i.e., the magnetic coupling across the vdW gap. The sign and strength of IMC have been found to vary significantly in different vdW magnets, and a comprehensive understanding on the underlying mechanism is still lacking.

The MnBi$_2$Te$_4$-family compounds as vdW layered intrinsic magnetic topological materials are particularly interesting for the rich magnetic topological phases they host, including the antiferromagnetic (AFM) topological insulator (TI) (*23-34, 39*) and the magnetic Weyl semimetal (WSM) (*23-26, 34*) in the 3D bulk, as well as the quantum anomalous Hall (QAH) insulator (*23-29, 33, 34*) and axion insulator (*23-28*) in their 2D films. Quantized transports have been observed in MnBi$_2$Te$_4$ thin flakes at relatively high temperatures (>1K) (*28, 29, 33*). Crystallized in a rhombohedral layered structure, bulk tetradymite-type MnBi$_2$Te$_4$ is composed of vdW-bonded septuple-layers (SLs). Each SL is consisted of seven atomic layers, of triangular lattice with ABC-stacking (*23-25*) in the sequence of Te/Bi/Te/Mn/Te/Bi/Te (Fig. 1(a)). The intralayer magnetic coupling is ferromagnetic (FM) with an out-of-plane easy magnetic axis, while the IMC is AFM. Thus the MnBi$_2$Te$_4$ bulk has an A-type AFM ground state.

The MnBi$_2$Te$_4$-family materials provide an excellent platform to explore the IMC mechanism of vdW magnets. Firstly, the ternary-compound family includes many candidate materials constituted by different combinations of nonmagnetic and magnetic elements (*23*), which enables a systematic study on the element-dependence of IMC. Secondly, the relatively long distance between neighboring magnetic atomic layers locks the way of bonds forming and forbids many kinds of electron hopping channels (see a more detailed analysis in Supplemental Material), simplifying the discussions on the IMC mechanism. Besides, since the topological properties of MnBi$_2$Te$_4$-family materials depend critically on the interlayer magnetic state (*23-25, 28, 33, 34*), the knowledge on the IMC mechanism can be directly used to engineer the magnetic topological quantum states and effects (*35-39*).

In this *Letter*, we systematically studied the constituent-element-dependence of the IMC of MnBi$_2$Te$_3$-family materials and heterostructures by first-principles calculations. Our results reveal that the IMC is dominated by the long-distance superexchange mechanisms mediated by the $p$ orbitals of non-magnetic atoms. Remarkably, both the sign and strength of IMC can be significantly manipulated by choosing different magnetic elements with different numbers of $d$ electrons and nonmagnetic elements of different electronegativities. This provides us guidelines to understand, design, and engineer the magnetic and topological properties not only for this material family but also for layered magnetic insulators in general.

First-principles calculations were performed in the framework of density functional theory (DFT) using the Vienna *ab initio* simulation package (VASP) (*40*). The Perdew-Burke-Ernzerhof (PBE) type exchange-correlation functional in the generalized gradient approximation (GGA) (*41*) was adopted. The localized 3$d$-states of V, Cr, Mn, Fe, Ni and 4$f$-states of Eu were treated by employing the PBE+$U$ approach ($U$ = 3.0, 4.0, 4.0, 4.0, 4.0, 5.0, respectively). The $U$ values have been tested and used by previous works (*7, 11, 19-25, 34*). Band structures for thin films and bulks were computed by the PBE+$U$ method and the modified Becke-Johnson (mBJ) functional

(*42*), respectively. VdW corrections were included by the DFT-D3 method (*43*). All the crystal structures were fully relaxed. The magnetic ground state and exchange energies for each material were determined by computing energy differences between different magnetic configurations. Topological surface states or edge states were calculated by using WannierTools package based on maximally localized Wannier functions (*44*). More calculation details are described in the Supplemental Material.

According to the superexchange theory (*45, 46*), the exchange coupling between two magnetic atoms emerges when an indirect electron hopping between $d$ (or $f$) orbitals of magnetic elements is mediated by the connecting $p$ orbitals. In the MnBi$_2$Te$_4$-family compounds, magnetic atoms are bonded with six chalcogen atoms in a slightly distorted octahedral geometry, where the $d$ orbitals of each spin state are roughly split to triply degenerate $t_{2g}$ states and doubly degenerate $e_g$ states. Generally, there exist three kinds of hopping channels in the system, including $e_g$-$p$-$e_g$, $e_g$-$p$-$t_{2g}$ and $t_{2g}$-$p$-$t_{2g}$, as illustrated in Fig. 1(c). Here σ bonds are formed between $e_g$ and $p$ orbitals, whose strength is much stronger than that of π bonds formed between $t_{2g}$ and $p$ orbitals. Therefore, the $e_g$-$p$-$e_g$ hopping channels, if existing, play a dominant role in determining the sign and strength of IMC.

We first analyze the IMC of MnBi$_2$Te$_4$. For the Mn ions, their $3d$ orbitals of one spin (spin-up) channel are fully occupied, and the other spin channel is empty. While the interlayer electron hopping between $d$ orbitals of neighboring Mn atoms is prohibited in the FM configuration, it gets allowed in the AFM configuration (Fig. 1(d)). Therefore, the IMC of MnBi$_2$Te$_4$ is AFM, giving an A-type AFM ground state as learned from previous theoretical and experimental works (*23-28, 30-32*). Further analyses on other MnBi$_2$Te$_4$-family compounds indicate that their IMC is always AFM, independent of constituent elements, as confirmed by our first-principles calculations. This seems to be a disadvantage for research and applications, since sometimes a FM IMC is desired, for instance, for obtaining a non-zero net magnetism and for realizing the QAH effect.

In contrast to pure compounds, vdW heterostructures of XBi$_2$Te$_4$ (X=Mn, V, Ni, Eu) has been rarely explored, which could open new opportunities to tuning magnetic and topological properties. In particular, we will show that the IMC can be driven into FM in the vdW heterostructures with different magnetic elements. We take the MnBi$_2$Te$_4$-EuBi$_2$Te$_4$ (MBT-EBT) combination as an example. The schematic lattice structure of the MBT-EBT bilayer is shown in Fig. 1(b). In EuBi$_2$Te$_4$, the 4$f$ orbitals of Eu are far away from the Fermi level ($E_F$), leaving the 5$d$ orbitals close to $E_F$. These 5$d$ orbitals are empty and polarized by magnetic moments of the 4$f$ electrons, giving spin-up states lower than spin-down ones (*47*). Although the $e_g$-$p$-$e_g$ hopping is allowed in both FM and AFM IMC configurations of MnBi$_2$Te$_4$-EuBi$_2$Te$_4$, the energy difference between the spin-up $e_g$ orbitals of Mn and Eu atoms is smaller in the former case, making the FM coupling energetically more favorable than the AFM one (Fig. 1(e)). Therefore, the IMC between a MnBi$_2$Te$_4$ SL and an EuBi$_2$Te$_4$ SL is FM (see a detailed analysis in the Supplemental Material). Applying similar analyses to all XBi$_2$Te$_4$-YBi$_2$Te$_4$ heterostructures, we find that generally the FM IMC emerges when X has $d$ electrons ⩾ 5 while Y has $d$ electrons < 5. In some complicated cases, such as MnBi$_2$Te$_4$-VBi$_2$Te$_4$, the FM and AFM IMC are comparable in strength, giving rise to relatively weak IMC.

The IMC mechanism is supported by our first-principles calculations. The energy differences between AFM and FM configurations per unit cell ($E_{ex}$) of a series of XBi$_2$Te$_4$-YBi$_2$Te$_4$ bilayers (X, Y = Mn, V, Ni, and Eu) are listed in TABLE 1. It is shown that MnBi$_2$Te$_4$-EuBi$_2$Te$_4$, MnBi$_2$Te$_4$-VBi$_2$Te$_4$, NiBi$_2$Te$_4$-EuBi$_2$Te$_4$, and NiBi$_2$Te$_4$-VBi$_2$Te$_4$ favor FM IMC, and the other combinations favor AFM IMC. The results can be understood by the superexchange theory as discussed above.

Noticeably, the superexchange coupling in magnetic oxides is supposed to be rather short-ranged (*45, 46*). Typically, only the cation-anion-cation coupling is considered. But the calculated IMC strength between MnBi$_2$Te$_4$-family materials is not negligible though the magnetic atoms from neighboring SLs are spaced by six layers of non-magnetic atoms together with a vdW gap. Especially, |$E_{ex}$| between two NiBi$_2$Te$_4$ SLs is as high as 16.5 meV. According to the traditional superexchange theory (*45, 46*), the IMC strength is mainly dependent on the hopping term between magnetic atoms, which is determined by the overlapping of the relevant orbitals in real space. In magnetic oxides, the electronegativity difference between oxygen ($\chi = 3.44$) and magnetic metal atoms (mostly $\chi < 2.00$) is so large that the medial $p$ electrons are rather localized around each oxygen ion. In the MnBi$_2$Te$_4$-family materials, the similar electronegativities between cations (Bi) and anions (Te) lead to much more delocalized $p$ electrons, which may facilitate longer-range magnetic coupling.

To verify the argument, we calculated the dependence of $|E_{ex}|$ on the non-magnetic elements in Ni(V)$_2$(VI)$_4$ compounds, where (V)= Bi or Sb, and (VI)=Te, Se, or S. As shown in Fig. 2(a), replacing Te ($\chi = 2.10$) with Se ($\chi = 2.55$) and S ($\chi = 2.58$) significantly reduces $|E_{ex}|$, from 16.5meV to 3.5meV and 1.1meV, respectively. The underlying reason is that Te $5p$ electrons are more delocalized than Se $4p$ and S $3p$ electrons in this system, and thus easier to hop to neighboring atoms along the $p$ orbital chain. In contrast, $|E_{ex}|$ drops to nearly zero ($< 0.1$meV) if artificially replacing Te with O, which is expected by the typical superexchange coupling for such a large interatomic distance. A similar trend is also found in the NiSb$_2$(VI)$_4$ compounds. Replacing Bi ($\chi =2.02$) with Sb ($\chi =2.05$) in NiBi$_2$(VI)$_4$ enhances $|E_{ex}|$ to as large as 24.7 meV in the NiSb$_2$Te$_4$ bilayer. The relationship of $|E_{ex}|$ and the electronegativity difference ($\Delta\chi$) of constituent elements are plotted for Ni(V)$_2$(VI)$_4$ compounds in Fig. 2(b), which shows a clear inverse correlation.

To further confirm this argument, we analyzed element-projection band structures of bulk NiSb$_2$Te$_4$, NiBi$_2$Te$_4$, NiSb$_2$Se$_4$, and NiBi$_2$Se$_4$ along the Γ-Z direction (out-of-plane) (Fig. 2(c)). The wider band dispersion along the Γ-Z direction means the more delocalized $p$ or $d$ electrons along out-of-plane direction. Comparing NiSb$_2$Te$_4$ with NiBi$_2$Te$_4$, the dispersion of Te bands in the former case is obviously wider than in the latter, facilitated by a closer electronegativity between Sb and Te. More hybridized and coupled with Te $5p$ bands, Sb $5p$ bands are pushed upward by Te $5p$ bands almost above Ni $3d$ bands, with wider band dispersion than that of Bi $6p$ bands. These band features will definitely lead to larger IMC in NiSb$_2$Te$_4$ than NiBi$_2$Te$_4$. Comparing NiSb$_2$Te$_4$ with NiSb$_2$Se$_4$, however, two factors are needed to be considered. Firstly, Te shares closer electronegativity to Sb than Se, and Sb $5p$ bands are located nearer to Te $5p$ bands than

Se 4$p$ bands. The dispersion of Sb 5$p$ bands in NiSb$_2$Te$_4$ is thus larger than NiSb$_2$Se$_4$, indicating more delocalized $p$ electrons. Secondly, Te 5$p$ orbitals are more extended than Se 4$p$ orbitals, leading to stronger overlapping with Ni 3$d$ orbitals and with Te 5$p$ orbitals of neighboring SLs in real space. Figure 2(d) shows the zoom-in band structures around Ni 3$d$ bands. The 3$d$ band dispersion of magnetic atoms is directly related to the IMC strength. Clearly, the dispersions of the four compounds give band widths NiSb$_2$Te$_4$>NiBi$_2$Te$_4$> NiSb$_2$Se$_4$>NiBi$_2$Se$_4$, exhibiting the same trend of IMC strength. The dependence of |$E_{\text{ex}}$| on magnetic elements listed in TABLE I can also be understood with similar analyses. The much stronger IMC of NiBi$_2$Te$_4$ bilayer than other XBi$_2$Te$_4$-YBi$_2$Te$_4$ combinations is because the 3$d$ bands of Ni is closer to and thus more strongly hybridized with $p$ bands than the $d$ bands of other magnetic elements.

The understanding on the IMC mechanism of MnBi$_2$Te$_4$-family enables us designing and engineering topological states by controlling the IMC. For example, bulk MBT is an AFM TI in its ground state and becomes a magnetic Weyl semimetal when it is changed into the FM configuration by a large external magnetic field (*23-25, 27*). The magnetic Weyl semimetal possesses only one pair of Weyl points and is expected to show high-Chern-number QAH phases in thin films (*33, 48-52*). The FM IMC between MBT and EBT suggests that one may obtain an intrinsic ferromagnetic WSM (in no need of applying external magnetic fields) in MBE-EBT heterostructures.

Figure 3(a) displays the calculated bulk band structure of a MBT-EBT superlattice. A band crossing is observed at $E_F$ near the Γ point along the Γ − Z direction. Except for the two crossing points (the other point is along opposite direction symmetrically) and a weakly dispersed band connecting them, no electron or hole pocket exist at $E_F$. Figure 3(b) gives the Fermi surface of the ($k_x - k_z$) surface states, showing a Fermi arc connecting the two crossing points (i.e. Weyl points). The band and Fermi surface structures indicate that bulk MBT-EBT is a type-I ferromagnetic WSM possessing only one pair of Weyl points (WPs) (*53*). By checking the motions of the sum of Wannier charge centers (WCCs), the two WPs show opposite chiralities, as theoretically expected (Fig. 3(c)). We checked the evolution of its band structure with the SOC artificially varied from 75% to 110% of the real strength. The dependence of band gap at the Γ point on the SOC strength is depicted in Fig. 3(d). The band gap closes twice at 92% and 102%, and the band closing points divide the phase diagram into three parts, including a trivial FM insulator phase, a WSM phase, and a FM insulator phase with band inversion. The last one is a phase named as "FMTI", which is topologically equivalent to Cr- or V-doped (Bi,Sb)$_2$Te$_3$ TIs (*54, 55*). When the time reversal symmetry is recovered above the Curie temperature, it is a typical TI. Below the Curie temperature, its bulk phase becomes topologically trivial with broken time-reversal symmetry, but its thin films are expected to be QAH insulators with *C*=1 (*51*).

A 2D thin film of a magnetic WSM can show the QAH effect with the Chern number increasing with the film thickness (*33, 51, 52*). Figure 4(a) shows the evolution of the Chern number and band gap of MBT-EBT thin films with varying film thickness. The film including one MBT-EBT bilayer is a normal insulator with a zero Chern number due to quantum confinement effects. Thicker films containing 2 to 10 MBT-EBT bilayers are QAH insulators with *C*=1, whose band gap reaches a maximum (about tens of meV) in the 3-bilayer film. In a magnetic WSM film, the increment of Chern number (Δ*C*) with film thickness satisfies (*33*): $\Delta C \approx \Delta N |k'_W|$, where $|k'_W|$ denotes the ratio of $\Gamma - W$ to $\Gamma - Z$ along $\Gamma - Z$ direction. For bulk MBT-EBT, $|k'_W| = \frac{0.0717}{0.5} \approx \frac{1}{7}$, which is consistent with the calculation result that *C*=2 from 11 to 17 bilayers and *C*=3 above 18 bilayers. The edge states of a 3-bilayer film and a 12-bilayer film are displayed in Fig. 4(b) and Fig. 4(c), respectively. Clearly, they are QAH insulators with *C*=1 and *C*=2, respectively, showing one and two gapless chiral edge modes crossing the bulk gap.

In this material system, a moderate external hydrostatic pressure can lead to a band inversion at the Γ point, resulting in a topological phase transition (*51*). Figures. 3(e) and 3(f) present the projection band (p-band) results projected by Bi (blue) and Te (red) under ambient pressure and 0.2 GPa hydrostatic pressure, respectively. Under ambient pressure, there is only one band inversion between Bi $p_z$ and Te $p_z$ bands, which leads to a WSM phase. Under 0.2 GPa pressure, two band inversions occur, opening an insulating band gap. The so-called "FMTI" thus emerges (*51*). Thin films of this phase above 2 bilayers are all QAH insulator with $C=1$, as confirmed by edge-state results shown for 3 bilayers (Fig. 4(e)) and 12 bilayers (Fig. 4(f)).

The IMC mechanism of MBT-family materials revealed in this work provides us a universal framework for understanding the magnetism of insulating layered magnetic compounds and their heterostructures. Below we summarize several guiding principles for estimating the IMC of insulating layered materials qualitatively or even quantitively.

First, the rules for analyzing superexchange interactions in ferrites can basically also be applied to estimate the sign and relative magnitude of IMC. Especially, $e_g$-$e_g$ hopping through $p$ bands usually dominates the magnetic coupling if present. The IMC of many vdW heterostructures such as $XBi_2Te_4$-$CrI_3$, $XBi_2Te_4$-$Cr_2Ge_2Te_6$, $XBi_2Te_4$-$FePS_3$ can be estimated by this way (see Supplemental Material). For systems where $e_g$-$e_g$ is not the dominated exchange coupling channel, more careful analyses are needed because different hopping channels favoring different kinds of magnetic states may compete (*7-10, 12-18*).

Second, a stronger IMC requires more delocalized medial $p$ electrons which can be realized by similar electronegativities between cations and anions. It's worth mentioning that heavy nonmetallic elements share the similar electronegativities with heavy metallic elements. One can choose materials including heavier elements near the metal-insulator boundary to obtain a stronger IMC.

Third, the IMC can be further enhanced if the $d$ bands of magnetic ions are near $E_F$, strongly hybridizing with the $p$ bands. So, $Ni^{2+}$-based FM or AFM insulators, for example, could contribute to a strong IMC with a neighboring layered magnetic material.

These guiding principles enable us designing and engineering the magnetic properties of magnetic vdW materials for different purposes. The various van der Waals heterostructures composed of different layered magnetic materials will make the principles easily tested and widely applied in experiment.

In conclusion, we systematically studied the IMC mechanism of the $MnBi_2Te_4$-family compounds and heterostructures. The superexchange coupling is found to dominate the IMC, and the delocalized $p$ electrons contribute to the relatively strong coupling strength across a long atomic distance. The understanding on the IMC mechanism implicates the principles of manipulating the interlayer magnetic configurations of insulating vdW magnets, which is especially valuable in exploring novel magnetic topological phases and effects in them.

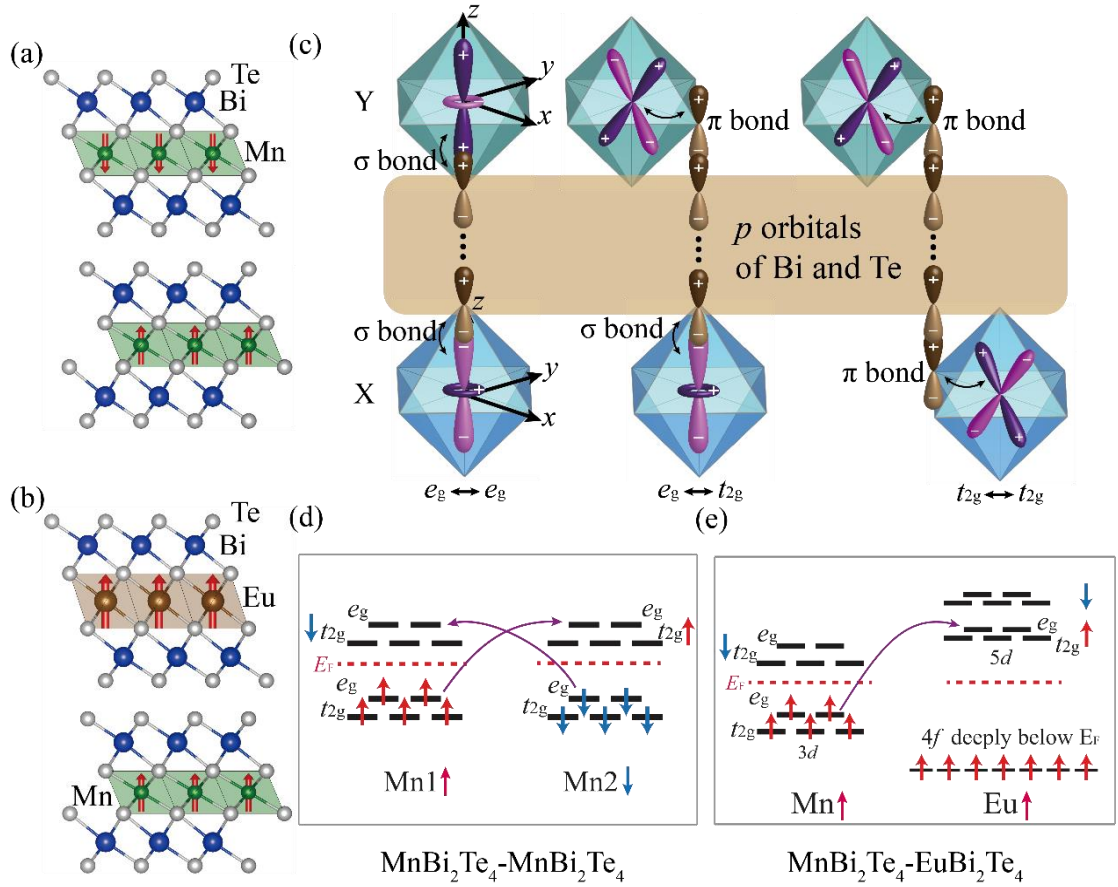

**Figure. 1.** (a, b) Schematic lattice structures (side view) of (a) a $MnBi_2Te_4$-$MnBi_2Te_4$ bilayer and (b) a $MnBi_2Te_4$-$EuBi_2Te_4$ bilayer. The red arrows indicate the spin directions. (c) Schmatic drawings of three hopping channels supporting the superexchange interlayer magnetic coupling: $e_g$-$p$-$e_g$ (left), $e_g$-$p$-$t_{2g}$ (middle) and $t_{2g}$-$p$-$t_{2g}$ (right). X and Y denote two magnetic atoms in $XBi_2Te_3$-$YBi_2Te_3$. (d, e) Schemtic drawings of the energy levels of magnetic atoms and the hopping channels between them in $MnBi_2Te_3$-$MnBi_2Te_3$ (d) and in $MnBi_2Te_3$-$EuBi_2Te_3$.

**TABLE 1:** The magnetic ground states and the interlayer exchange energies ($E_{ex}$) which is the energy difference between the A-type AFM state and FM state of $XBi_2Te_3$-$YBi_2Te_3$ bilayers (X, Y represents Ni, Mn, V, or Eu).

| $E_{ex}$/(meV) | | X | | | |
|---|---|---|---|---|---|
| | | $Ni(3d^8)$ | $Mn(3d^5)$ | $V(3d^3)$ | $Eu(4f^7+5d^0)$ |
| Y | $Ni(3d^8)$ | AFM -16.5 | AFM -4.6 | FM +3.5 | FM +1.1 |
| | $Mn(3d^5)$ | AFM -4.6 | AFM -1.3 | FM +1.2 | FM +0.6 |
| | $V(3d^3)$ | FM +3.5 | FM +1.2 | AFM -0.5 | AFM -0.2 |
| | $Eu(4f^7+5d^0)$ | FM +1.1 | FM +0.6 | AFM -0.2 | AFM -0.1 |

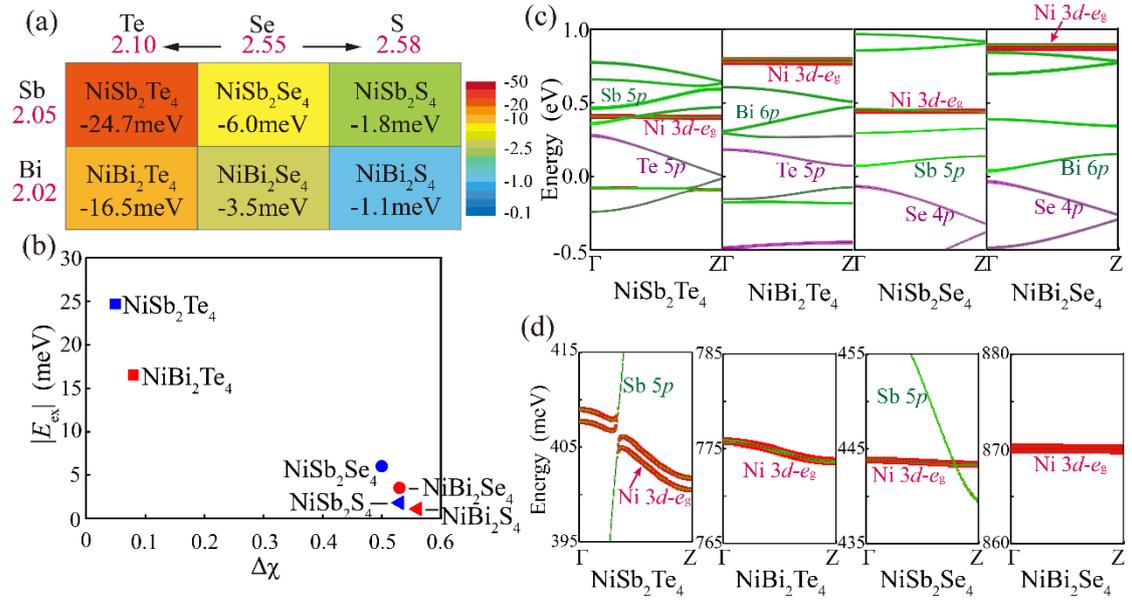

**Figure. 2.** (a) Table of the interlayer magnetic coupling (IMC) strength ($|E_{ex}|$) of six compounds of Ni(V)$_2$(VI)$_4$, in which (V) = Bi, Sb; (VI) = Te, Se, S. Electronegativity ($\chi$) of each element is noted below as red digits. (b) Dependence of $|E_{ex}|$ on the electronegativity difference ($\Delta\chi$) between the anions (VI) and nonmagnetic cations (V) of six Ni(V)$_2$(VI)$_4$. (c) Dispersions of element-projected bands along $\Gamma - Z$ directions of NiSb$_2$Te$_4$, NiBi$_2$Te$_4$, NiSb$_2$Se$_4$, and NiBi$_2$Se$_4$. Red curves denote the Ni 3$d$-$e_g$ bands. Green and purple curves denote the $p$ bands of (V) and (VI), repectviely. (d) Zoom-in figures near the Ni 3$d$-$e_g$ bands in the four materials.

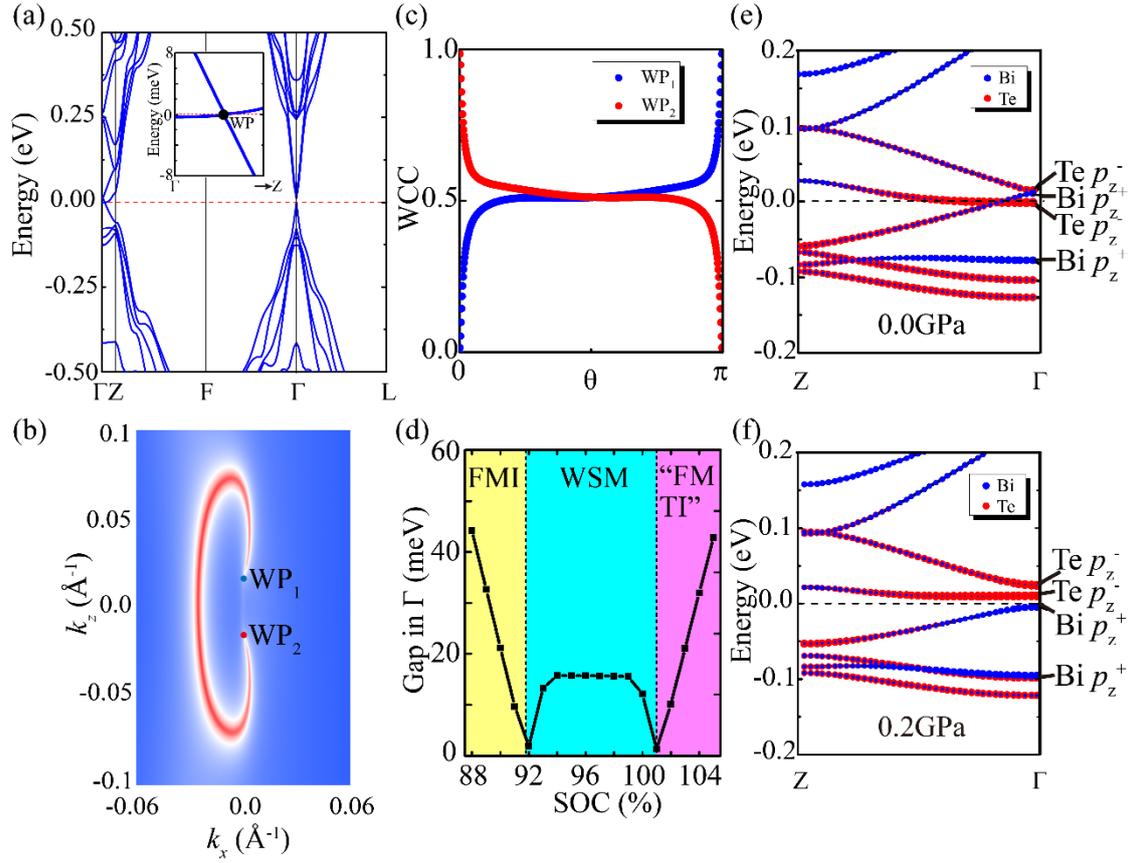

**Figure. 3.** Band structres and topological features of bulk MnBi$_2$Te$_4$-EuBi$_2$Te$_4$. (a) Band structures of bulk MnBi$_2$Te$_4$-EuBi$_2$Te$_4$ superlattice. The inset shows a weakly dispersed band along the Γ-Z direction, implying the type-I WSM nature. (b) Surface states on the (110) plane at the energy of the two Weyl points (WP$_1$ and WP$_2$). (c) The motions of the sum of Wannier charge centers on a sphere surrounding each of two Weyl points in the momentum space. (d) Band gap at Γ as a function of the strength of spin-orbit coupling. The three areas of different colors denote different topological phases. (e, f) Element-projected bandstructures of Bi $p_z$ and Te $p_z$ orbitals along Z − Γ direction under (e) zero pressure and (f) external 0.2GPa hydrostatic pressure, respectively. "+" and "−" signs denote even and odd parities, respectively.

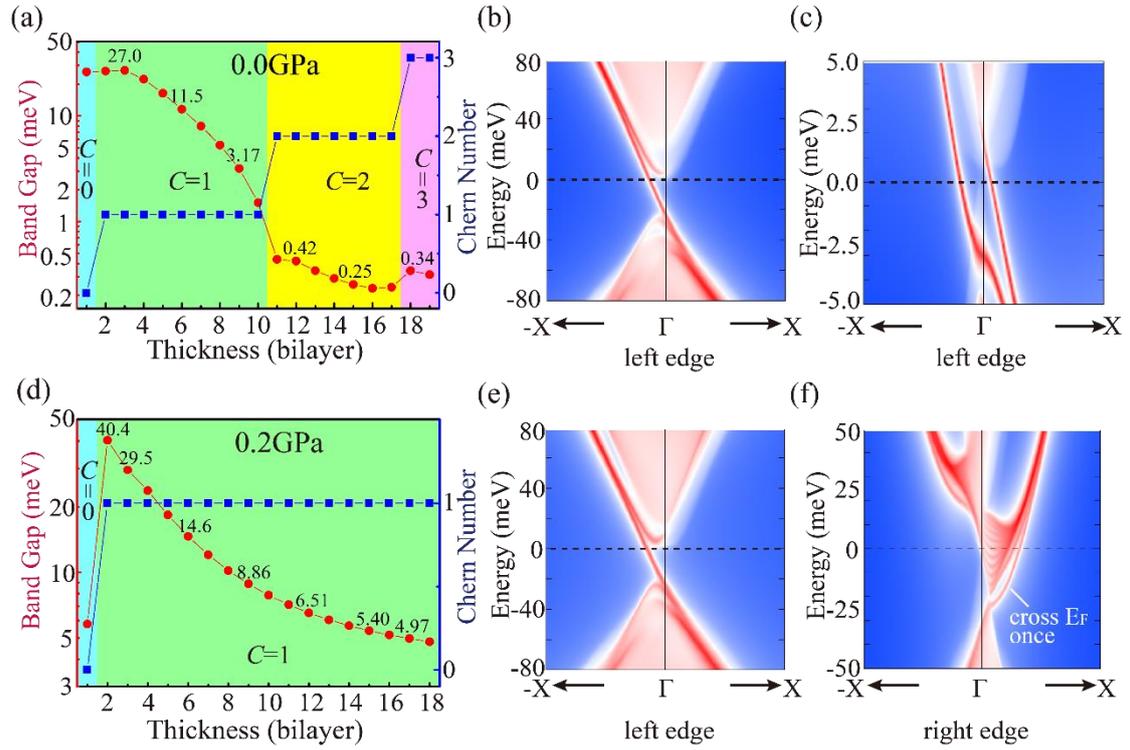

**Figure. 4.** Topological features of MnBi$_2$Te$_4$-EuBi$_2$Te$_4$ superlattice thin films. (a, d) The band gap size at Γ point (red) and the Chern number (blue) of MnBi$_2$Te$_4$-EuBi$_2$Te$_4$ thin film with increasing film thickness under hydrostatic pressue of (a) zero and (d) 0.2 GPa. The areas of different colors represent the phases with different Chern numbers. (b, c) Edge states of (b) a film of 3 MnBi$_2$Te$_4$-EuBi$_2$Te$_4$ bilayers and (c) a film of 12 MnBi$_2$Te$_4$-EuBi$_2$Te$_4$ bilayers without external pressure, which show one and two chiral edge states, respectively. (c, e) Edge states of (e) a film of 3 MnBi$_2$Te$_4$-EuBi$_2$Te$_4$ bilayers and (f) a film of 12 MnBi$_2$Te$_4$-EuBi$_2$Te$_4$ bilayers under 0.2GPa, both of which exhibit one chiral edge state.